\documentclass[apjl]{emulateapj}
\usepackage{graphicx}
\usepackage{amssymb}
\usepackage{amsmath}
\usepackage{natbib}

\def\Chandra{\textit{Chandra}\:}

\shorttitle{High-redshift X-ray Jet With No Radio Counterpart}
\shortauthors{A. Simionescu et al.}

\begin{document}

\title{Serendipitous discovery of an extended X-ray jet without a radio counterpart\\ in a high-redshift quasar}

\author{A.~Simionescu\altaffilmark{1}, {\L}.~Stawarz\altaffilmark{2}, Y.~Ichinohe\altaffilmark{1,\,3}, C.~C.~Cheung\altaffilmark{4}, M.~Jamrozy\altaffilmark{2}, A.~Siemiginowska\altaffilmark{5}, K.~Hagino\altaffilmark{1}, P.~Gandhi\altaffilmark{6}, and N.~Werner\altaffilmark{7,\,8}}
\affil{$^1$Institute of Space and Astronautical Science (ISAS), JAXA, 3-1-1 Yoshinodai, Chuo-ku, Sagamihara, Kanagawa, 252-5210 Japan}
\affil{$^2$Astronomical Observatory, Jagiellonian University, ul. Orla 171 , 30-244 Krak\'ow, Poland}
\affil{$^3$Department of Physics, Graduate School of Science, University of Tokyo, 7-3-1 Hongo, Bunkyo, Tokyo 113-0033, Japan}
\affil{$^4$Space Science Division, Naval Research Laboratory, Washington, DC 20375-5352, USA}
\affil{$^5$Harvard-Smithsonian Center for Astrophysics, 60 Garden St., Cambridge, MA 02138, USA}
\affil{$^6$School of Physics \& Astronomy, University of Southampton, Hampshire SO17 1BJ, Southampton, United Kingdom}
\affil{$^7$KIPAC, Stanford University, 452 Lomita Mall, Stanford, CA 94305, USA}
\affil{$^8$Department of Physics, Stanford University, 382 Via Pueblo Mall, Stanford, CA  94305-4060, USA}

\begin{abstract}
A recent \Chandra observation of the nearby galaxy cluster Abell\,585 has led to the discovery of an extended X-ray jet associated with the high-redshift background quasar B3\,0727+409, a luminous radio source at redshift $z=2.5$. This is one of only few examples of high-redshift X-ray jets known to date. It has a clear extension of about $12''$, corresponding to a projected length of $\sim100$\,kpc, with a possible hot spot located $35''$ from the quasar. The archival high resolution VLA maps surprisingly reveal no extended jet emission, except for one knot about $1.4''$  from the quasar. The high X-ray to radio luminosity ratio for this source appears consistent with the $\propto(1+z)^{4}$ amplification expected from the inverse Compton radiative model. This serendipitous discovery may signal the existence of an entire population of similar systems with bright X-ray and faint radio jets at high redshift, a selection bias which must be accounted for when drawing any conclusions about the redshift evolution of jet properties and indeed about the cosmological evolution of supermassive black holes and active galactic nuclei in general.

\end{abstract}

\keywords{galaxies: active --- galaxies: jets --- quasars: individual (B3\,0727+409) --- radiation mechanisms: non-thermal --- radio continuum: galaxies --- X-rays: general}

\section{Introduction}\label{intro}

Over the years, the \Chandra X-ray Observatory has revealed a significant number of X-ray bright, kiloparsec-scale jets in active galactic nuclei \citep[AGN; e.g.,][]{Harris2006}\footnote{\texttt{http://hea-www.harvard.edu/XJET/}}. Despite notable progress in understanding these objects, it is still unclear what is the composition of the plasma carrying the energy, how and where the jet particles are accelerated, and how relativistic the jets are in terms of their bulk velocity, $\beta$, and Doppler beaming factors, $\delta\equiv[\Gamma(1-\beta\cos\theta)]^{-1}$, with $\theta$ being the jet inclination and $\Gamma\equiv(1-\beta^{2})^{-\frac{1}{2}}$ the jet bulk Lorentz factor.

Radiative models for the broad-band emission of large-scale quasar jets have also remained a matter of debate. The two main candidate mechanisms for producing the jet X-ray emission are synchrotron and inverse Compton scattering of the cosmic microwave background (IC/CMB). A robust determination of the jet energetics requires constraints on the relative importance of these two processes. The IC/CMB scenario implies particle-dominated and highly relativistic outflows ($\delta \simeq 10$), which do not suffer severe deceleration or energy dissipation between sub-pc and kpc scales \citep[e.g.,][]{Ghisellini2001,Tavecchio2007}, while the synchrotron interpretation can be reconciled with highly-magnetized and possibly slower jets on kpc distances from the quasar cores \citep[e.g.,][]{Stawarz2004,Hardcastle2006}. 

The IC/CMB model predicts an increase in the X-ray--to--radio flux ratio with redshift, due to the amplification of the CMB energy density \citep[e.g.,][]{Schwartz2002,Ghisellini2014}; neglecting a weak dependance on the spectral slope of the non-thermal continuum, 
\begin{equation} 
\frac{[\nu F_{\nu}]_x}{[\nu F_{\nu}]_r} \propto \left(\frac{u'_{\rm CMB}}{u'_{\rm B}}\right)\,\left(\frac{\delta}{\Gamma}\right)^2 \propto(1+z)^{4}\,\left(\frac{\delta}{B'}\right)^2\, ,
\end{equation}
where $F_{\nu}$ is the observed energy flux spectral density, $u'_{\rm CMB}\simeq4\times10^{-13}\,\Gamma^2\,(1+z)^4$\,erg\,cm$^{-3}$ is the CMB energy density in the jet rest frame (denoted hereafter by primes), and $u'_{\rm B} \equiv {B'}^2/8 \pi$ is the comoving energy density of the jet magnetic field. Studying high-redshift jets can thus provide important clues on the mechanism responsible for their observed X-ray emission. However, very few high-redshift X-ray jets are known to date \citep{Siemiginowska2003, Yuan2003, Cheung2006, Cheung2012}. Therefore, variations of the jet beaming, $\delta$, and jet magnetic field strength, $B$, both from system to system and between different knots along the same jet, can introduce a scatter that may mask the expected $(1+z)^{4}$ increase even when the IC/CMB model is the dominant emission mechanism \citep[see][]{Cheung2004}. 

Increasing the sample of high-redshift quasar jets with good-quality X-ray data will not only allow us to distinguish between competing radiative models; by combining in-depth studies of the cosmological evolution of the jet properties from the epoch of the quasar formation up to the present day with our current understanding of black hole growth and the evolution of black hole spin \citep[e.g.,][]{Volonteri2013}, we can shed new light on the jet launching mechanism, as well as on the evolution of the intergalactic medium that the jets propagate through.

Here, we report the discovery of an extended, X-ray bright, radio faint black hole jet associated with the quasar B3\,0727+409. Based on the Sloan Digital Sky Survey (SDSS) Data Release 9 \citep{ahn1}, the spectroscopic redshift of this source is $z=2.500\pm0.001$. Assuming a $\Lambda$CDM cosmology with $\Omega_{\Lambda}=0.73$, $\Omega_{\rm M}=0.27$, and $H_{\rm 0}=71$\,km\,s$^{-1}$\,Mpc$^{-1}$, the redshift of B3\,0727+409 corresponds to the luminosity distance of $d_{\rm L}\simeq20.1$\,Gpc and the conversion scale $\simeq8$ kpc/$\arcsec$.

\section{Observations and Data Reduction}

\subsection{\Chandra}

The X-ray jet associated with B3\,0727+409 was discovered in a relatively short (20~ks) \Chandra observation from 2014 December 15 (ObsID 17167), targeting the nearby galaxy cluster Abell\,585 \citep[$z\simeq0.121$; see][]{Jamrozy2014}. The brightest cluster galaxy of Abell\,585 is located $\sim2.5'$ northwest of the quasar, so that both were observed in the $8.3'\times8.3'$ field-of-view of the Advanced CCD Imaging Spectrometer (ACIS) S3 chip.

We reprocessed the standard level 1 event lists produced by the \Chandra pipeline in the standard manner, using the CIAO (version 4.7) software package to include the appropriate gain maps and updated calibration products. Bad pixels were removed and standard grade selections applied. The information available in VFAINT mode was used to improve the rejection of cosmic ray events. Periods of anomalously high background were excluded by examining the light-curve of the observation in the $0.3-10$\,keV energy band using the standard time binning methods recommended by the \Chandra X-ray Center. The net exposure time after cleaning is 19\,ks. 

\subsection{VLA}

Inspecting the Very Large Array (VLA) images of B3\,0727+409 previously presented in \citet{Jamrozy2014} that probe a range of spatial scales $>1\arcsec$, we did not find any significant radio emission associated with the X-ray jet. We therefore analyzed an additional set of VLA data from the NRAO\footnote{The National Radio Astronomy Observatory is a facility of the National Science Foundation operated under cooperative agreement by Associated Universities, Inc.} archive from 2007 August 6, when it was observed as a calibrator in program AL696. The observations were collected in the most extended A-configuration and consisted of five short scans centered at 1.43\,GHz and three scans at 4.86\,GHz. The data were calibrated with AIPS \citep{bri94} following standard procedures with amplitude calibration utilizing a scan of 3C\,286 at 1.43\,GHz and 3C\,147 at 4.86\,GHz. After editing, the total exposure times for B3\,0727+409 were approximately 9 and 4 minutes at the respective frequencies. The ($u,v$) data were imported into DIFMAP \citep{she94} for phase and amplitude self-calibration. CLEAN images were convolved with circular Gaussians with full-width half maxima matched to the natural weighted beam sizes of $0.4''$ at 4.86\,GHz and $1.5''$ at 1.43\,GHz (0.09 and 0.14 mJy beam$^{-1}$ off-source rms, respectively). We also reimaged the VLA 4.71 GHz B-array dataset from \citet{Jamrozy2014} to provide an image with the same $1.5''$ beam as the 1.43 GHz image.

\begin{figure}[!t]
\includegraphics[width=0.5\textwidth]{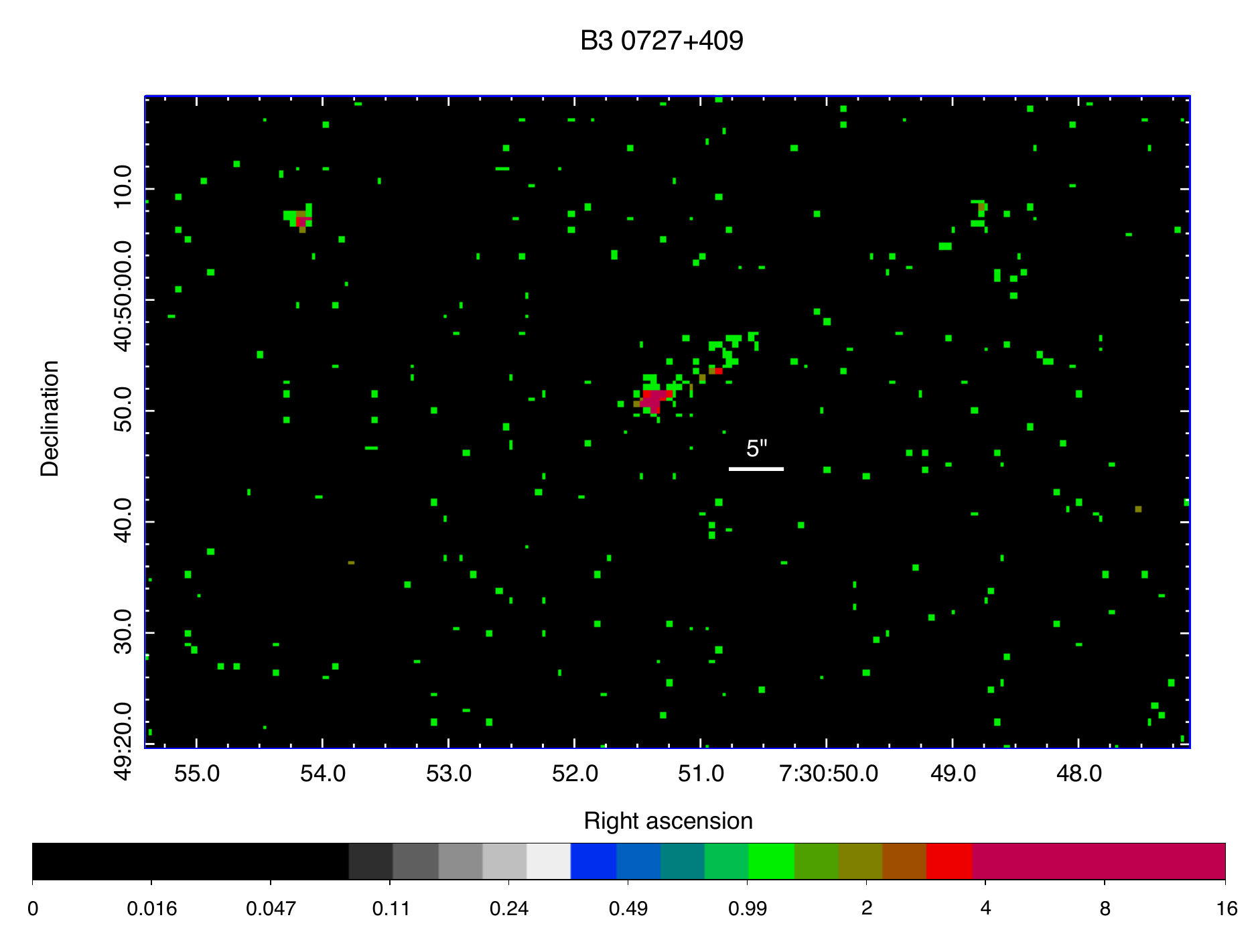}
\caption{Unsmoothed \Chandra count map showing the B3\,0727+409 quasar core and X-ray jet. Colorbar units are counts per native ACIS pixel ($0.492''\times0.492''$).}
\vspace{0.1cm}
\label{imgX}
\end{figure}

\section{Results}

\subsection{Imaging}

Figure\,\ref{imgX} shows a raw \Chandra count map of the region around B3\,0727+409 in the $0.6-7.5$\,keV energy band. Because the jet is relatively bright and compact, no vignetting or background corrections were applied. The image reveals a clear extension of $\sim12''$ northwest of the quasar's core, corresponding to a projected length of $\sim100$\,kpc at $z=2.5$. Note that there is no visible charge-transfer streak in the ACIS-S3 read-out direction (offset from the X-ray jet by $\sim 60^\circ$ clockwise) that would indicate significant pile-up of counts from the bright core.

In our 4.86\,GHz VLA $0.4''$ resolution map, we confirm the detection of a radio feature $\sim1.4''$ from the quasar core found by \citet{Gobeille2014} using the same data. We find no significant additional emission coinciding with the rest of the X-ray jet. Figure\,\ref{imgXR} shows a \Chandra image of the jet, rebinned to 0.1 ACIS pixels and smoothed with a $0.5''$ Gaussian filter, overlaid with 4.86\,GHz radio contours. 

Note that very long baseline interferometry maps show apparent superluminal motion up to $\sim6.6c$ in a one-sided milliarcsecond-scale jet oriented toward the extended structure seen in X-rays, indicating relativistic beaming on small scales \citep{Britzen2008}.

\subsection{Spectral fitting and flux measurements}

\subsubsection{Extended jet}

For the position on the detector corresponding to the quasar core, the 90\% enclosed-counts fraction aperture of the \Chandra point-spread function (PSF) has a radius of $r_{PSF90}=1.77''$. We extracted X-ray spectra from a $10''$-long rectangular region with a width of $2\,r_{PSF90}$, starting at a minimum distance of $r_{PSF90}$ and extending out to $10''+r_{PSF90} \simeq 12''$ from the quasar core.

We fit the resulting spectrum in the 0.6--7.5 keV energy range with an absorbed power-law model, with the hydrogen column density fixed to the Galactic value, $N_{\rm H}=6.2\times 10^{20}$\,cm$^{-2}$ \citep{Kalberla2005}, assuming no intrinsic absorption. The background was estimated from an annulus centered on the X-ray peak, with inner and outer radii of $15''$ and $30''$. The best-fit power-law index is $\Gamma_x=1.74^{+0.34}_{-0.32}$, with a corresponding flux density at 1~keV of $2.7\pm0.7$ nJy. 

This extraction region does not include the radio knot at $\sim1.4''$, which is only marginally resolved from the quasar core with \Chandra (Section \ref{sect_knot}). The extended jet is thus remarkably radio faint, with no visible emission in both VLA images. To estimate upper limits, we used the $1.5''$-beam images and defined four adjacent $1.5\arcsec\times2\arcsec$ apertures elongated along the X-ray jet direction (position angle $PA=-60^\circ$), centered 3.1\arcsec, 5.4\arcsec, 7.7\arcsec, and 10.0\arcsec\ from the quasar. Using the AIPS task UVSUB to subtract the cores from the ($u,v$) data, as well as the modeled Gaussian representing the $1.4''$ knot in the 1.43 GHz data (below), the $3\sigma$ point source upper limits in each of the respective apertures were $<0.6$, $<0.6$, $<0.6$, and $<0.4$\,mJy at 1.43\,GHz, and $<0.8$ (larger due to contamination from the adjacent $1.4\arcsec$ knot), $<0.3$, $<0.4$, and $<0.3$\,mJy at 4.71\,GHz. Assuming the radio jet is composed of a series of four unresolved radio knots within the defined apertures, the total radio emission co-spatial with the X-ray jet visible beyond the $1.4''$ knot is thus $<2.2$ and $<1.8$ mJy, corresponding to $[\nu F_{\nu}]_x/[\nu F_{\nu}]_r>205$ and $>73$ at 1.43 and 4.71\,GHz, respectively.

\begin{figure}[!t]
\includegraphics[width=0.5\textwidth]{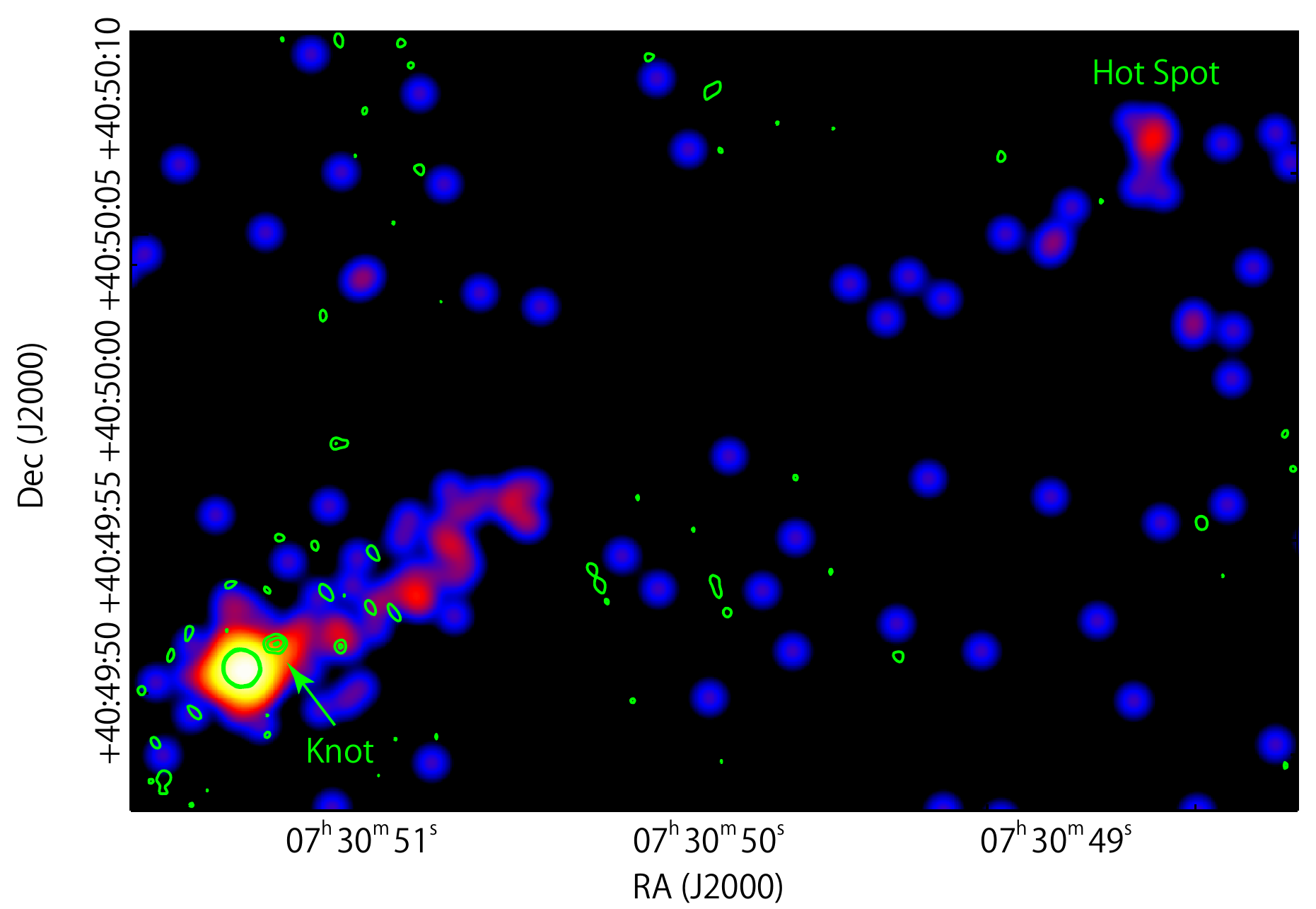}
\caption{A zoom-in on the X-ray image of B3\,0727+409 shown in Figure\,\ref{imgX}, smoothed with a $0.5''$ Gaussian filter. VLA 4.86\,GHz contours at 0.29, 0.58, and 0.87 mJy beam$^{-1}$ are shown in green. We have corrected for a slight offset between the X-ray and radio contours due to the \Chandra astrometric error.}
\label{imgXR}
\end{figure}

In addition, we detect excess X-ray emission (a total of 10 counts) located $35''$ from the quasar core (280 kpc, projected), along the same position angle as the jet (Figure \ref{imgXR}). The hypothesis that this excess of counts is solely due to Poisson fluctuations around the average background level estimated from a neighbouring region is ruled out at the $p$-value of $2.18\times10^{-5}$; since it lacks any SDSS counterpart, this X-ray feature may therefore likely represent the terminal hotspot of the B3\,0727+409 jet. The corresponding $3\sigma$ point source upper limits at the location of this X-ray hotspot are $<0.6$ mJy (1.43 GHz) and $<0.3$ mJy (4.71 GHz).

\subsubsection{Jet knot at $1.4''$}\label{sect_knot}

By measuring the count rates in four partial annuli with opening angles of $90^\circ$ and spanning $0.5-1\,r_{PSF90}$ from the quasar core, we estimate that the azimuth corresponding to the radio knot contains $6.3\pm3.5$ X-ray counts above the surface brightness level determined from the other three azimuths. Assuming this emission follows the same power-law index as the extended jet, this implies the knot's flux density at 1~keV is $0.44\pm0.10$ nJy, and the 2--10 keV luminosity of the entire jet (summing the knot and extended jet) is $\sim(6.1\pm2.5)\times10^{44}$\,erg\,s$^{-1}$. 

From the VLA data, we measured a flux density for the knot of 1.5\,mJy at 4.86\,GHz by fitting a circular Gaussian component in the ($u,v$) plane with the {\tt modelfit} program in DIFMAP. Though at the edge of the beam of the bright core in the lower-resolution 1.43\,GHz image, that same knot is clearly visible in the CLEAN components, and we measured a flux density of 4.5\,mJy. In the presence of the bright core, we estimate the uncertainties in the flux densities are $20\%$ at both frequencies. The resultant spectral index of the knot is $\alpha=0.90\pm0.23$. The broad-band emission from the knot therefore corresponds to the flux density ratio $[\nu F_{\nu}]_x/[\nu F_{\nu}]_r\simeq15$, with very similar values at both frequencies.

\begin{table*}
\caption{Flux measurements for the B3\,0727+409 jet and quasar core. Upper limits are quoted at the $3\sigma$ level. }
\begin{center}
\begin{tabular}{ccccc}
\hline
\hline
 & core & $1.4''$ knot & extended jet & hot spot \\
\hline 
\Chandra counts (0.6--7.5 keV) & $212\pm15$ & $6.3\pm3.5$ & $38\pm6$ & $10\pm3$ \\ 
X-ray power-law index $\Gamma$ & $1.32\pm0.12$ & 1.74 (assumed) & $1.74^{+0.34}_{-0.32}$ & 1.74 (assumed) \\
$0.6-7.5$\,keV flux (erg\,cm$^{-2}$\,s$^{-1}$) & $1.26_{-0.13}^{+0.2}\times10^{-13}$ & $\sim0.32\times10^{-14}$ & $1.92_{-0.62}^{+0.50}\times10^{-14}$ & $\sim0.51\times10^{-14}$ \\ 
2--10 keV luminosity (erg\,s$^{-1}$) & $2.6_{-0.5}^{+0.4}\times10^{45}$ & $\sim0.9\times10^{44}$ & $5.2^{+1.9}_{-3.0}\times10^{44}$ & $\sim1.4\times10^{44}$ \\
flux density at 1\,keV (nJy) & $11.4\pm1.2$ & $0.44\pm0.10$ & $2.7\pm0.7$ & $0.71\pm0.18$ \\
\hline
VLA 1.43\,GHz flux density (mJy) & $294\pm15$ & $4.5\pm0.9$ & $<2.2$ & $<0.6$ \\
VLA 4.86\,GHz flux density (mJy) & $243\pm12$ & $1.5\pm0.3$ & -- & -- \\
VLA 4.71\,GHz upper limit (mJy) & -- & -- & $<1.8$ & $<0.3$ \\
\hline 
\end{tabular}
\end{center}
\label{tab_flux}
\end{table*}

\subsubsection{Quasar core}

We also extracted spectra from a circular region of radius $r_{PSF90}$ centered around the quasar core; subtracting the local background as described above and fitting the resulting spectrum with an absorbed power-law, we obtain a best-fit index $\Gamma_x=1.32\pm0.12$ and a 2--10 keV rest frame luminosity of $L_{X,core}=2.6_{-0.5}^{+0.4}\times10^{45}$ erg\,s$^{-1}$. No intrinsic absorption in addition to the Galactic $N_{\rm H}$ is required by the fit. 

All our X-ray and radio flux measurements are summarised in Table \ref{tab_flux}.

\section{Discussion}\label{disc}

\subsection{Origin of the jet X-ray emission}

A non-thermal origin provides the most natural explanation for the jet X-ray emission. We modeled the radio--to--X-ray spectrum of the large-scale jet in B3\,0727+409 with the IC/CMB scenario assuming a broken power-law shape of the electron energy distribution, $n_e\!(\gamma')\propto{\gamma'}^{-p}$ for $\gamma'_{min}\leq\gamma'\leq\gamma'_{br}$, and $n_e\!(\gamma')\propto{\gamma'}^{-p-1}\, \exp[-\gamma'/\gamma'_{max}]$ for $\gamma'>\gamma'_{br}$, where $\gamma'$ is the electron Lorentz factor. This parametrization takes into account the expected break in the electron spectrum resulting from radiative cooling. We also assume pressure equipartition between the emitting electrons and the jet magnetic field, and allow for a heavy jet content with one $e^-p^+$ pair per 1--10 $e^{\pm}$ pairs \citep[see][]{Sikora2000}. Finally, we assume a spherical geometry for the $1.4''$ knot with a $3$\,kpc radius (consistent with the knot's radio extent in the VLA 4.86\,GHz image), and a cylindrical geometry for the extended jet with the same radius and a length of $80$\,kpc.

\begin{figure}[!t]
\includegraphics[width=0.5\textwidth]{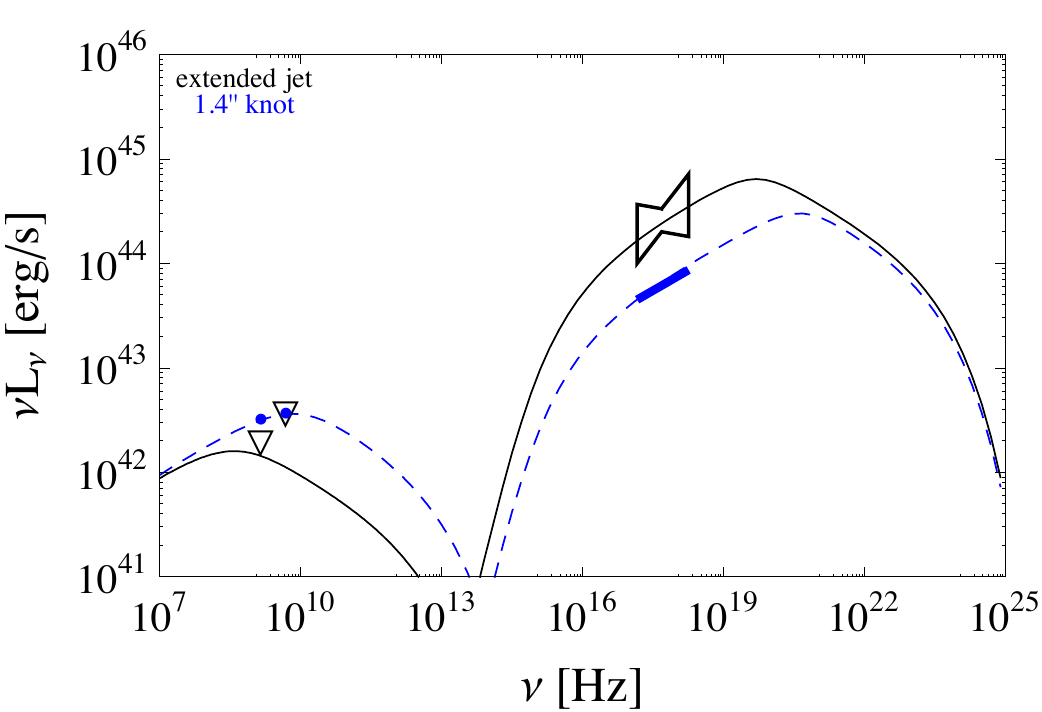}
\caption{The spectral energy distributions of the B3\,0727+409 extended jet and $1.4''$ knot, along with the corresponding synchrotron plus IC/CMB model curves.}
\label{sed}
\end{figure}

The broad-band jet spectrum can be described by the model shown in Figure \ref{sed}, returning reasonable parameters: the jet inclination $\theta\simeq7.5^\circ$, the jet bulk Lorentz factor $\Gamma\simeq10$ (meaning $\delta\simeq7.4$), the jet magnetic field decreasing slowly along the outflow from $B'\simeq30$\,$\mu$G down to $15$\,$\mu$G, a ``standard'' form of the electron energy distribution with $\gamma'_{min}=10$, $\gamma'_{max}=10^5$, $p=2.5$, and the cooling break decreasing from $\gamma'_{br}=3\times10^3$ at the position of the $1.4''$ knot down to $\gamma_{br}=10^3$ at the position of the outer jet.

Although all parameters are currently only weakly constrained by the data, this model seems to favour a high total jet kinetic power of the order of $L_j\sim(0.3-3)\times 10^{47}$\,erg\,s$^{-1}$ (depending on the exact proton content), and highly relativistic jet bulk velocities maintained over very large scales of order the de-projected source size, $\ell_{dep}\gtrsim600$\,kpc. However, the estimated jet kinetic power further depends on various assumptions in our model: for a purely leptonic jet, $L_j$ would be smaller than cited here, while if the magnetic field energy density is below the equipartition value, $L_j$ would increase, while the required relativistic beaming would decrease. Note also that, alternatively, a synchrotron origin of the X-ray emission is not excluded by the data (however, see the discussion at the end of Section \ref{cosmo}).

On the other hand, a thermal origin of the X-ray emission from the extended jet region is implausible: the best-fit thermal model (employing the {\tt apec} model with a metallicity fixed at 0.2 Solar) implies a best-fit temperature of $kT=17.8_{-6.8}^{+33.9}$\,keV, which would be unusually high for any truly diffuse structure at $z=2.5$. Also the corresponding electron density, $n_e$, and total mass would be exceptional: assuming the same cylindrical geometry as above, the best-fit emission measure of the {\tt apec} model translates to $n_e\sim0.8\,f^{-1/2}$\,cm$^{-3}$ and $M_{gas}\sim4\,f^{1/2}\times10^{10}\,M_{\odot}$, with $f<1$ denoting the filling factor of the gas (with respect to the assumed cylindrical geometry). Such large amounts of very hot gas aligned with radio jets have not been observed in luminous quasars, although see \citet{Carilli2002} for the $z=2.2$ radio galaxy PKS\,1138-262.

\subsection{Cosmological context}\label{cosmo}

In the past, several claims have been made for distinct jet properties of high-redshift quasars when compared with their low-redshift analogs. For example, \citet{Volonteri2011} argued for a decrease of the average bulk Lorentz factor of high-redshift jets, because of an apparent deficit of luminous SDSS radio-loud quasars above $z\sim3$ with respect to the model predictions based on the Swift/BAT (Burst Alert Telescope) hard X-ray survey. Additionally, \citet{Singal13} demonstrated that the ``radio-loudness'' (i.e., the ratio of the 5\,GHz core radio flux to the $B$-band core flux) of the SDSS$\times$FIRST (Faint Images of the Radio Sky at Twenty cm) quasar population increases with redshift.

B3\,0727+409 appears particularly interesting in this context for several reasons. First, the X-ray luminosity of its large-scale jet is only 5--6 times lower than the X-ray luminosity of the core. This is in contrast to the lower-redshift quasars targeted with \Chandra, for which the jet--to--core X-ray luminosity ratio is typically $\sim0.01$, or lower \citep[see][]{Marshall2005}. Thus, the active nucleus of B3\,0727+409 seems \emph{under-luminous} in X-rays with respect to the emission of its large-scale outflow. 

Second, the unresolved core of B3\,0727+409 appears particularly radio-loud for its accretion rate. Using the SDSS spectrum, \citet{Jamrozy2014} estimate the mass of the central black hole to be between $M_{\rm BH}=(3.33\pm1.70)\times10^{8}M_{\odot}$ (from the MgII line) and $(2.26\pm0.34)\times10^{8}M_{\odot}$ (based on the CIV line). The bolometric luminosity of the accreting matter estimated from the MgII line is $L_d\simeq1.5\times10^{45}$\,erg\,s$^{-1}$, meaning the accretion rate is $\dot{m}_{acc}\simeq\eta_d^{-1}\,(L_d/L_{Edd})\sim0.4$ in Eddington units, for the standard $\eta_d\simeq10\%$ radiative efficiency of the accretion disk. 
Given this rate, B3\,0727+409 appears to be characterized by a surprisingly large radio-loudness parameter of $\mathcal{R}\simeq10^6$, at least 100 times larger than local quasars with comparable $L_d/L_{Edd}$ ratios \citep[see][]{Sikora2007}. Note that our IC/CMB modeling implies moreover $L_j/L_{Edd}\sim1-10$, which is consistent with a very high (maximum) efficiency of the jet production in high-$z$ sources \citep{Tchekhovskoy2011}.

\begin{figure}[!t]
\includegraphics[width=0.45\textwidth]{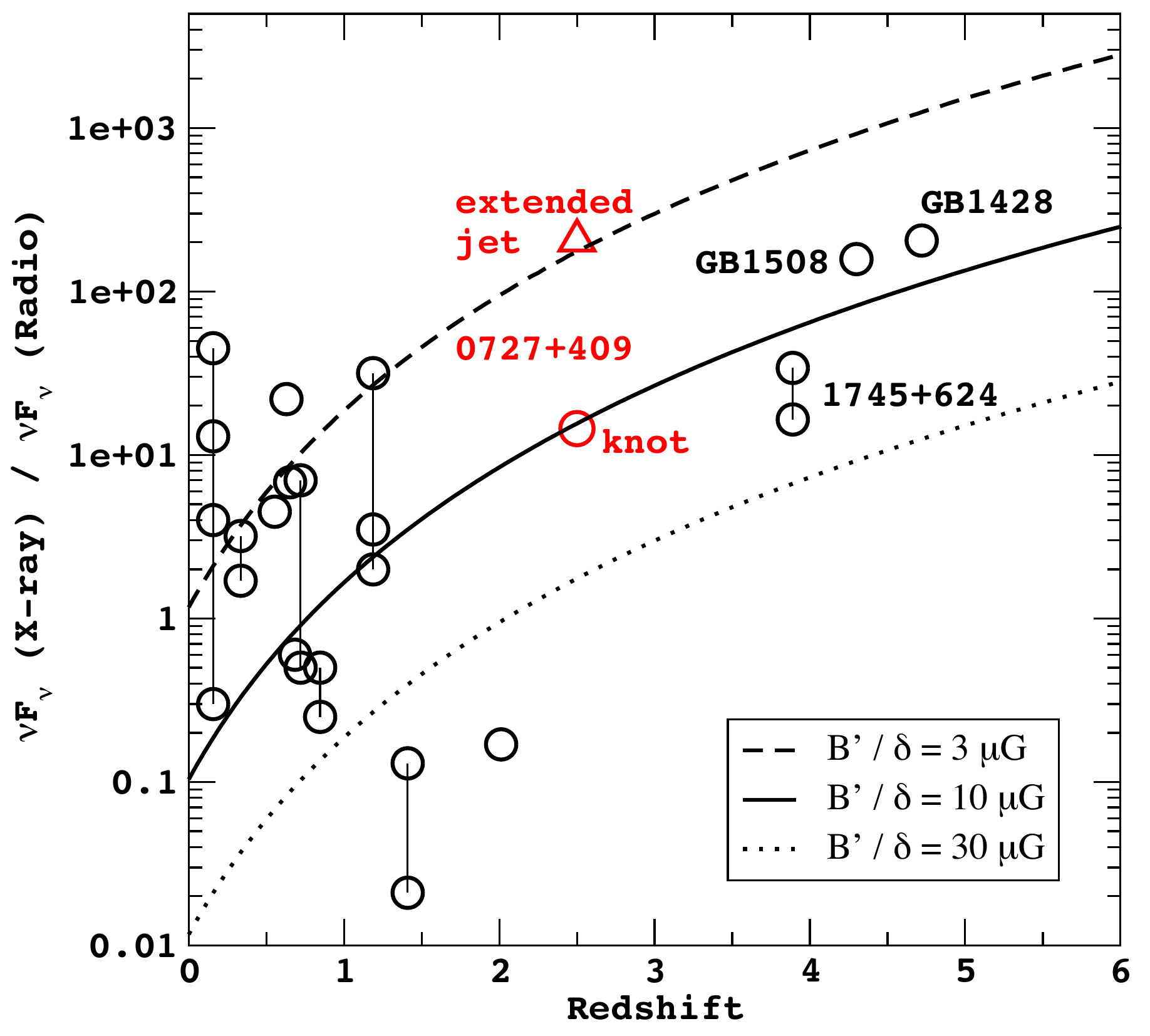}
\caption{Plot of the $[\nu F_{\nu}]_x/[\nu F_{\nu}]_r$ ratio vs. redshift for X-ray quasar jets detected with \Chandra \citep[adapted from][with the addition of two subsequently detected $z>3.5$ examples]{Cheung2004}. B3\,0727+409 is shown in red; the triangle represents the 3$\sigma$ lower limit for the part of the jet lacking radio detection. Curves indicate the expected energy flux ratio in the framework of the IC/CMB scenario for given combinations of $B'$ and $\delta$ (see Equation\,1). Different knots from the same jet are connected by thin vertical lines.}
\label{nufnu}
\end{figure}

Finally, in Figure \ref{nufnu}, we compare the measurements of $[\nu F_{\nu}]_{x(1\:keV)}/[\nu F_{\nu}]_{r(1.4\:GHz)}$ for B3\,0727+409 to those of previously published large-scale X-ray jets detected with \Chandra. The X-ray--to--radio luminosity ratios for our target appear consistent with the amplification expected from the IC/CMB model in the case of large jet bulk velocities. This seems to disfavour a synchrotron origin of the jet X-ray emission in B3\,0727+409. Nonetheless, the statistical uncertainties are large due to the relatively short exposure time, and the redshift distribution of \Chandra jets at $z\gtrsim2-3$ is still rather sparsely sampled. Deeper X-ray and radio data will provide significantly improved constraints on the jet emission model for this quasar.

\section{Summary}

We report on the serendipitous discovery of an extended ($\sim12''$, or $\sim100$\,kpc projected) X-ray jet associated with the $z=2.5$ quasar B3\,0727+409, for which the archival VLA maps reveal no radio counterpart (except for a single knot $\sim1.4''$ from the quasar core). A possible X-ray hot spot is identified $35''$ (280\,kpc) from the quasar. The remarkable X-ray luminosity of the structure, $L_{\rm 2-10\,keV}\gtrsim6\times10^{44}$\,erg\,$s^{-1}$, implies a very efficient production of non-thermal X-ray photons by ultrarelativistic jet electrons. The large X-ray--to--radio luminosity ratio supports the scenario in which this is the inverse-Comptonization of the CMB photons which dominates radiative outputs of large-scale jets in the X-ray domain \citep[at least in ``core-dominated quasars'', since in the cases of sources observed at larger jet viewing angles, i.e. ``lobe-dominated quasars'' and FR\,II radio galaxies, the situation may be more complex; see, e.g.,][]{Kataoka2008,Cara2013,Gentry2015}. If correct, this would further imply a highly relativistic bulk velocity ($\Gamma\sim10$) maintained on hundreds-of-kpc scales, even at high redshifts, and a very high efficiency of the jet production ($L_j/L_{Edd}\gtrsim1$) already during the epoch of quasar formation.

The serendipitous discovery of such an object may signal the existence of an entire population of similar systems with bright X-ray and faint radio jets at high redshift, which will have been missed by the current observing strategies that mostly focus on \Chandra follow-up of known bright radio jets. Similar predictions for the ubiquity of X-ray bright, radio faint {\it lobes} at $z\geq2$ were put forward by e.g. \cite{Fabian2009} and \cite{Mocz2011}. The seemingly different properties of this source compared to local quasars suggest that this selection bias must be accounted for when drawing any conclusions about the redshift evolution of jet properties and indeed about the cosmological evolution of supermassive black holes in general.

\acknowledgments
\L.S. and M.J. were supported by Polish NSC grants DEC-2012/04/A/ST9/00083 and DEC-2013/09/B/ST9/00599, respectively. 
Y.I. acknowledges a Grant-in-Aid for Japan Society for the Promotion of Science (JSPS) Fellows. 
C.C.C. was supported at NRL by NASA DPR S-15633-Y. A.Sie. was supported by NASA contract NAS8-03060 to the \Chandra X-ray Center. N.W. acknowledges NASA grant GO5-16127X. 

{\it Facilities:} \facility{\Chandra}, \facility{VLA}.

\bibliographystyle{apj}

\end{document}